# Differences and Connections Between Individual (Leontief Type) Activities and Aggregate (Cobb-Douglas Type) Results

Carlos Esteban POSADA[1]

*Abstract*

*Each production establishment is assumed to have, at any given time, a unique combination of capital and labor (a Leontief function), but the aggregate output at that same time must still be modeled with a Cobb-Douglas function (or a CES, although the latter yields less efficiency). This has two implications: 1) the total factor productivity variable of the macroeconomic function is endogenous: It depends primarily on the technical factors of the individual establishments and, secondarily, on their levels of capital and labor.; 2) the optimization processes of any establishment cannot be instantaneous, even in the absence of (monetary) adjustment costs; they are processes occurring over several time stages and depending on expectations. However, these implications do not substantially contradict what would correspond to the optimization of a hypothetical firm described by a Cobb-Douglas (or CES) function.*

## I. Introduction

The output of an establishment (factory, farm, hydroelectric plant, store, etc.) at a given time is interpreted in this text as the result of an activity described (leaving aside the value of raw materials, etc.) by a Leontief function: a function with one characteristic: a unique combination of capital and labor. However, the aggregate output of all establishments is adequately represented by a Cobb-Douglas function or a *CES* function (with aggregated capital and labor in varying combinations), even though each individual production is subject to a Leontief function.

<u>The Micro Level</u>

For establishment *i* at moment *j*, the Leontief function is[2]:

---

[1] Professor, School of Finance, Economics, and Government; EAFIT University (Colombia). Email: cposad25@eafit.edu.co.
I thank the comments from José Miguel ARIAS, Cristian CASTRILLÓN, John Jairo GARCÍA, Álvaro HURTADO, and Alejandro TORRES on previous versions.
[2]This function is commonly presented in many Microeconomics texts, but Ferguson's book on production theory (1969) is worth highlighting.

$$(1) \quad Y_{i,j} = Min\left[\frac{l_{i,j}}{a_{i,j}}, \frac{k_{i,j}}{b_{i,j}}\right]; a_{i,j}, b_{i,j} > 0$$

Where $l_i$, and $k_i$ are the quantities of labor and capital, and $a_i$ , and $b_i$ are technical coefficients; their inverses are the average productivities of labor and capital, respectively.

## The Macro Level

With homogeneous productions and in competition (relative prices = 1), we have at the macro level:

$$(2) \qquad Y_j = \sum_{i=1}^{n} Y_{i,j}\,;\; L_j = \sum_{i=1}^{n} l_{i,j}\,;\; K_j = \sum_{i=1}^{n} k_{i,j}$$

And it is assumed that capital and labor are homogeneous.

With this document I seek to achieve three objectives: 1) Describe the output of the economy as a whole using a Cobb-Douglas (or alternatively it would be a Constant Elasticity of Substitution, *CES*) function based on individual productions generated by Leontief functions; 2) Take into account and explain the marginal productivities of capital and labor at the microeconomic level as conceived by the owner of an establishment or their representative when focusing on their case: producing efficiently (without waste) with a Leontief function; 3) Interpret the decisions of this agent as the rational ones approximating the profit-maximization process imposed by the agent of an "ideal" firm whose production function would be Cobb-Douglas or *CES*.

These objectives are pursued through the six following sections, namely: Sections II ("Individual Productions"), III ("Aggregate Production"), IV ("Cobb-Douglas Multifactor Productivity"), V ("Marginal Productivities with Leontief Functions and Approximation to Profit Maximization"), and VI ("Final Comments, and a Conclusion").

## II. Individual Productions

Suppose there are a relatively large number of establishments, each establishment i (i from 1 to n) operating at a given time *j* with a Leontief function. If we order them from lowest to highest according to their level of output, and considering that each one is characterized by a pair of coefficients $a_{i,j}, b_{i,j}$, and a pair of quantities of capital and labor, then, considering all the outputs, we will most likely find a positive nonlinear relationship between aggregate output per unit of labor and aggregate capital per unit of labor. Assuming 50 establishments and smooth, decreasing evolutions of both coefficients, their outputs (ordered from lowest to highest) are represented in Figure 1 as the broken line connecting points ABC (a possibility resulting from the condition on a minimum established by equation 1)[3].

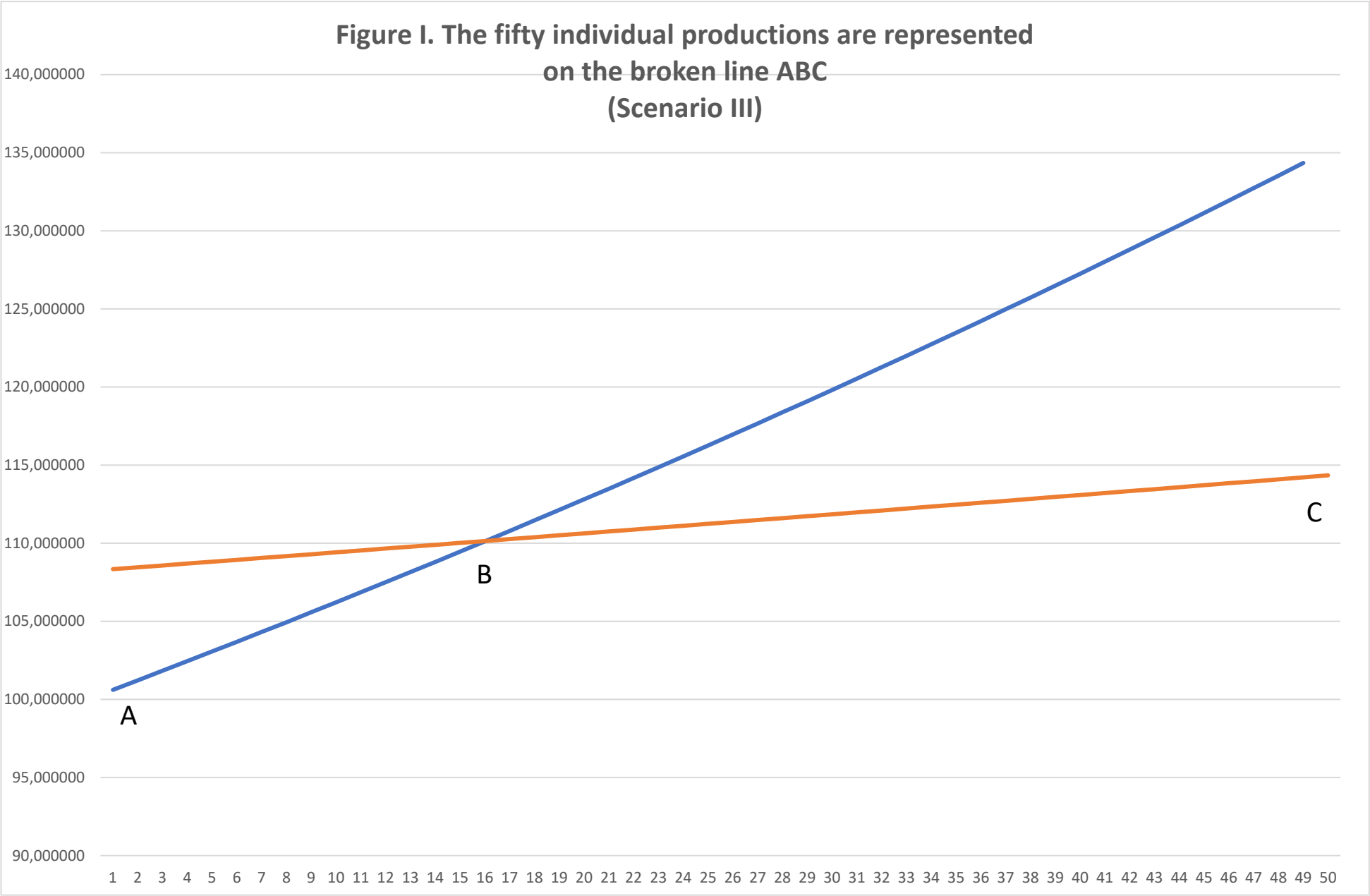

Figure 2 shows, for all 50 establishments, the relationships between output per unit of labor and capital per unit of labor for each establishment. The break derives from that observed in Figure 1.

[3] Graph *I* was constructed using figures from a scenario (scenario III) that assumes: 1) a single point in time, 2) the magnitudes of the coefficients *a* and *b* are smaller and the amount of capital per unit of labor of a firm is greater the higher its production, and 3) that there is a firm i (for *i* = 18) with a production level from which the output of this firm and of the firms that follow in ascending order ceases to be associated with $^{l\ i}/_{a_i}$ and becomes associated with $^{k_i}/_{b_i}$. This transition is what explains the break in the ABC line.

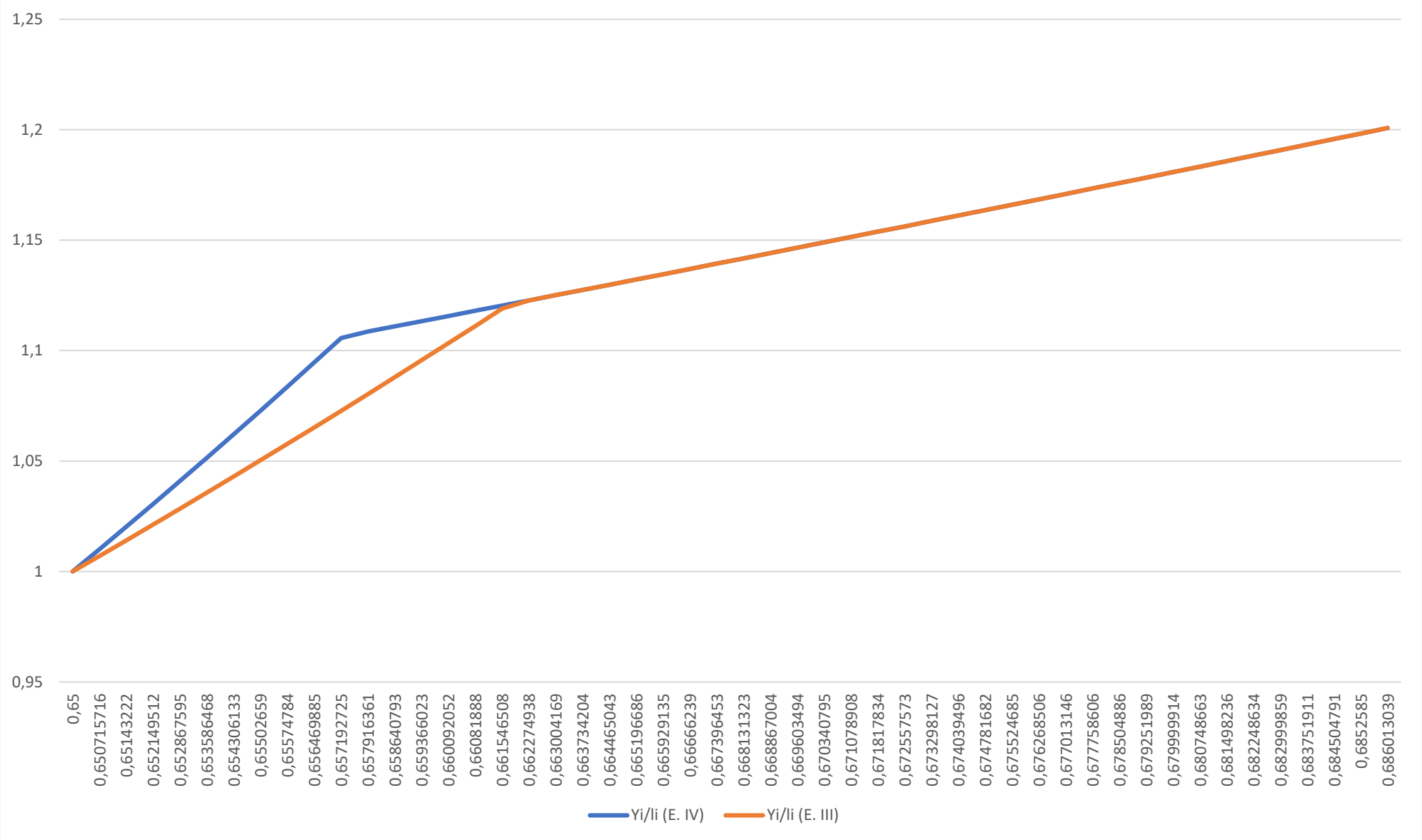

**Figure 2. Output/Labor (vertical axis) vs. Capital/Labor (horizontal axis) Scenarios III & IV**

The best fit to the broken line in Figure 2 is a second-degree equation. The relationship between the output/labor and capital/labor ratios that can be established based on the Cobb-Douglas function is a second-degree equation, as follows: with positive but decreasing slopes as the capital/labor ratio increases.

The previous figures correspond to two of four different scenarios designed for 50 establishments; in two scenarios breaks do not occur. But in the real world any economy has more than 50 establishments (and with high heterogeneity among them). Colombia has more than one million establishments. Thus, in any real-world economy the probability is practically zero that, if the set of establishments were ordered by production level at a given moment, at least one break like the one presented in Figure 1 would not be found, and therefore, at least one break like the one corresponding to Figure 2. And, in general, if, after the aforementioned ordering, thousands, hundreds of thousands, or millions of breakups were observed, the probability is zero that, in a representation of this analogous to that of Figure 2, the breakups with southward movements (in terms of Figure 2) would not be dominant, so that the statistical fit of the points would be represented as a second-degree equation, also with positive and declining slopes.

## III. Aggregate Production

Figure 2 serves to illustrate something that Jones (2005) and Growiec (2013) explained broadly and deeply, and that Houthakker (1955-56) proposed: the nonlinear aggregate production function with positive but decreasing marginal returns to its factors (the most common being Cobb-Douglas or CES) is a theoretical construct that can be conceived as relevant to describe a relationship between the economy's total output and its factors (in our case, aggregated capital and labor)[4], even if each individual production must be described by a Leontief function.

Based on the above, we can assume that an economy whose output is achieved by the activities of establishments operating under functions with fixed proportions of capital and labor (Leontief-type functions[5]) is an economy in which aggregate production maintains a Cobb-Douglas-type relationship with aggregate capital and labor, and under constant returns to scale, as follows:

$$(3)\ Y_j = Z_j K_j^{\alpha} L_j^{1-\alpha}$$

The following table shows this for four different scenarios, assuming the same 50 establishments and an elasticity of aggregate output to capital, α, equal to 0.5.

**Table 1. The Aggregate Production Function: Cobb-Douglas**

| Escenario | $K$ | $L$ | $K^{\alpha}$ | $L^{1-\alpha}$ | $Y$ | $Z = \frac{Y}{K^{\alpha}L^{1-\alpha}}$ |
|---|---|---|---|---|---|---|
| *I* | 3257.98 | 4879.44 | 57.08 | 69.85 | 4879.44 | 1.22 |
| *II* | 3250.00 | 5000.00 | 57.01 | 70.71 | 5491.08 | 1.36 |
| *III* | 3257.98 | 4879.44 | 57.08 | 69.85 | 5492.13 | 1.377 |
| *IV* | 3257.98 | 4879.44 | 57.08 | 69.85 | 5518.54 | 1.384 |

Scenario *I* was constructed assuming that technical coefficients $a_i$ , $b_i$ are equal in all establishments for every moment *j*, and that factor use intensities $\left(\frac{k_i}{l_i}\right)$ vary among establishments. Scenario *II* considers different coefficients among establishments, but factor use intensity is equal in all establishments. In Scenario *III*, both technical coefficients and factor use intensities differ. In Scenario *IV*, differences in coefficients among establishments are greater, but differences in factor use intensity are the same as in Scenario *III*.

## IV. Multifactor Productivity (or Total Factor Productivity) in the Cobb-Douglas Case

The results shown in Table 1 indicate that the main cause of the "jumps" observed in so-called total factor productivity Z among different scenarios is the change in technical coefficients (compare Scenario *III* with Scenario *I*), while the secondary cause is the change in factor use intensities (comparing Scenario *III* with Scenario *II*). And perhaps most importantly: the table makes it evident that Z is an endogenous variable although its magnitude could be partially influenced by some exogenous factor.

---

[4] Jones (2005) used the Pareto distribution (a probability distribution function) to generate values of the technical coefficient for hypothetical individual production levels that underpin an aggregate Cobb-Douglas production function. Growiec (2013) used another probability function, also for individual technical coefficients (the Weibull distribution) demonstrating that a *CES* production function could be considered a representation of aggregate production. However, for given levels of capital and labor, and for the same total factor productivity, the total output generated by the Cobb-Douglas function is higher than that generated by the *CES* function; that is, the *CES* function implies static inefficiency compared to the Cobb-Douglas function.

[5] This is without prejudice to accepting that in an establishment, in addition to its routine productive activity, tasks associated with projects for modifications to production processes may be carried out.

## V. Marginal Productivities, Factor Prices, and Approximation to Profit Maximization

What is expressed in this section will be better understood through numerical examples referring to any given establishment; among the 50, establishment 14 from Scenario II was randomly chosen, although the same exercise could have been done with any other establishment.

Since it is assumed that at any given moment *j* there is no possibility of contemporaneous changes in factor usage, the marginal productivity of each factor must be considered as the result of comparing the expectation formed by the agent at moment *j* about the output at *j+1* with the output at *j*, assuming that "tomorrow" (*j+1*) the agent will have (and use) one additional unit of the factor; all this under certain conditions that will be clarified in the following paragraphs.

Tables 2 and 3 show examples of marginal productivities of capital and labor, and decisions regarding the use of these factors. Once the tables are presented and the necessary clarifications made, it will become evident what must be assumed for the exercises to be logical and consistent with the concept of marginal productivity.

| Table 2. The *PmgK* (Marginal Productivity of Capital): An Expectation | | | | | | | |
|---|---|---|---|---|---|---|---|
| Moment | | $1/a_i$ | $1/b_i$ | $k_i/l_i$ | $k_i$ | $y_i$ | $Pmgk_i$ |
| $j$ | | 1.09562 | 1.68849 | 0.65 | 65 | 109.56 | Not available |
| $j+1$ | Expectation in *j* | 1.09649 | 1.68849 | 0.66 | 66 | 109.65 | 0.0872 |

| Table 3. The *PmgL* (Marginal Productivity of Labor): An Expectation | | | | | | | |
|---|---|---|---|---|---|---|---|
| Moment | | $1/a_i$ | $1/b_i$ | $k_i/l_i$ | $l_i$ | $y_i$ | $Pmgk_i$ |
| $j$ | | 1,09562 | 1,68849 | 0,65 | 100 | 109,56 | Not available |
| $j+1$ | Expectation in *j* | 1,09562 | 1,68868 | 0,644 | 101 | 109,76 | 0,202 |

According to Table 2, what would justify making "today" (moment *j*) the decision to marginally increase (by one unit) the capital "tomorrow" (*j+1*) would be the expectation of a higher level of average labor productivity "tomorrow" $\left(\frac{1}{a_{i,j+1}}\right)$, with everything else remaining constant. Indeed, the expected increase in output $(y_{i,j+1} - y_{i,j})$ would imply an expected marginal productivity of capital equal to 8.72% when capital increases by one unit, from 65 to 66 units.

Similarly, according to Table 3, the decision to marginally increase the amount of labor (comparing its magnitude tomorrow with today) would be justified by the expectation of a higher magnitude of the average capital productivity tomorrow $\left(\frac{1}{b_{i,j+1}}\right)$, with everything else remaining constant[6]. In this case, the marginal productivity of labor expected today for tomorrow is 0.202.

Formally, to generalize, the following applies to the hypothesis on the determination of the expected marginal productivity of capital by the expected average productivity of labor:

$$Z \equiv E_j a_{i,j+1}; Z\text{: } exogenous\ variable$$

$$Hypothesis\text{: } E_j a_{i,j+1} > a_{i,j}$$

$$Hypothesis\text{: } X = f(Z);\ f' > 0$$

$$Where\ \ X \equiv \left[\frac{E_j(Y_{i,j+1}) - Y_{i,j}}{E_j(k_{i,j+1}) - k_{i,j}}\right]$$

And it is postulated that the expected marginal productivity of labor is determined by the (expected) average productivity of capital. So, the following is postulated:

$$U \equiv E_j b_{i,j+1}; U\text{: } exogenous\ variable$$

$$Hypothesis\text{: } E_j b_{i,j+1} > b_{i,j}$$

$$Hypothesis\text{: } V = h(U); h' > 0$$

$$Where\ \ V \equiv \left[\frac{E_j(Y_{i,j+1}) - Y_{i,j}}{E_j(l_{i,j+1}) - l_{i,j}}\right]$$

One point must be emphasized: the appeal to the usual assumption regarding cross marginal productivities (example: marginal changes in the quantity of factor 1 imply changes in the same direction in the marginal productivity of factor 2). Furthermore, it is necessary to avoid circular reasoning; and circularity is easily avoided insofar as we consider the expectation about the average productivity of each factor for period *j+1* as exogenous. The basis for this consideration is the very definition of the Leontief function, which implies that the average productivity of each factor is the inverse of the technical coefficient accompanying it, according to equation 1, and such coefficient is exogenous.

---

[6] Chapters 2 and 3 of Ferguson (1969) include an analysis of the marginal productivities of different factors of production in Leontief functions. However, the analysis is static, and therefore the marginal productivity of a factor must be understood in terms of what would happen to the output (the difference in output, comparing alternatives at a given time) if there were a marginal difference in the magnitude of a factor (let's say factor x). In such an analysis, and under the assumption of constant technical coefficients and assuming that factor x is the limiting factor of output, marginal productivity is calculable, positive, and equal to average productivity of the factor x, and therefore is a constant. I believe the conclusion of all this is that such an analysis is a formal exercise lacking economic substance.

These considerations do not exclude the possibility of using alternative assumptions; however, in any case, it is useful—especially when dealing with Leontief functions—to distinguish between today's realities and the decisions made today based on what is expected today for tomorrow.

In any case, if expectations are confirmed, the changes will be permanent; otherwise, they will be reversed (starting from *j+2*).

If, in the face of expectations of increases in the average productivities of capital or labor, the quantities of the corresponding factors to be used are increased without regret "tomorrow," when the results are known, it can be said that the process replicates that of profit maximization in a firm with a Cobb-Douglas production function (taking into account, of course, that budget constraints must be respected).

Thus, new expectations regarding average productivities of capital or labor create gaps between expectations about marginal productivities of labor or capital and the prices of one or the other factor (real wage or real interest rate). If these expectations are confirmed, in subsequent periods (*j+1, j+2*, …) the factor (or both) will be increased until the decline in its marginal productivity closes the gap. And, of course, the fall in the marginal productivity of a factor also implies a fall in its average productivity.

What is stated in this section means that the process of adjusting the magnitudes of the production factors to their optimal levels is slow; that is, it requires at least two periods, even under the implicit assumption that there are no monetary costs of adjusting capital or labor to their optimal levels. But if such costs existed, adjustments could be even slower.

Moreover, variations in the marginal productivities of capital or labor tomorrow will cause changes in marginal cost, comparing its magnitude tomorrow with today. Therefore, the assumption that all productive activities are adequately represented by Leontief functions is not an obstacle to describing changes in marginal costs; but these changes must be understood as events occurring over time (the succession of moments).

What has been said so far makes it necessary to clarify that the case of firms or establishments producing under the Leontief modality does not imply that their managers or shareholders are unable of reacting or indifferent to changes in the relationship between the cost of labor and that of capital. A change in such a relationship, if judged permanent, is perceived as an incentive to modify the proportion between the quantities of these factors, but the adjustment will not be instantaneous: at the present moment, the perception of the incentive will trigger a process of information gathering, project development, etc., leading to a change in the capital-to-labor ratio in one or several future periods. And this process is usually carried out in stages or gradually, so completing it may require several moments (and at each moment, each establishment will produce with a Leontief process).

An example could be an automotive company with several plants (establishments) and a project for an additional one. In such a case, the easiest (or least difficult or least costly) option may be to review the design of the new plant so that it meets the production conditions most suited to the expectation regarding the anticipated relationship between the real interest rate and the real wage.

## VI. Final Comments, and a Conclusion

What has been discussed here is relevant for understanding that changes in the so-called multifactor productivity and in the first and second derivatives of the aggregate production function are simplified expressions of structural changes at the microeconomic level, where agents operate within Leontief-type processes. These changes originate either from expectations and the pursuit of greater profits or from factors or processes unrelated to production theory.

Over time, the proportions between capital and labor change, albeit gradually, and this occurs due to new expectations regarding the productivities of the production factors (with technical change being fundamental) or due to changes in the price relationship between these factors. Rationality (for example, trying to avoid enormous costs) requires that optimization processes not be instantaneous; they must be gradual: slow over time.